\documentclass[a4paper]{jpconf}
\usepackage{graphicx}
\begin{document}
\title{High spin structures in the $A\approx 40$ mass region:
from superdeformation to extreme deformation and clusterization
(an example of $^{28}$Si)}

\author{A. V. Afanasjev and D. Ray }

\address{Department of Physics and Astronomy, Mississippi State
University, Mississippi State, MS 39762, USA}

\ead{Anatoli.Afanasjev@gmail.com}

\begin{abstract}
  The search for extremely deformed structures in the yrast and 
near-yrast region of $^{28}$Si has been performed within the cranked 
relativistic mean field theory up to spin $I=20\hbar$. The fingerprints 
of clusterization are seen (well pronounced) in the superdeformed 
(hyperdeformed) configurations.

\end{abstract}

\section{Introduction}

  A systematic search for extremely deformed structures in the 
$N=Z$ and $N=Z+2$ S, Ar, Ca, Ti and Cr nuclei has been performed
for the first time in the framework of covariant density functional
theory in \cite{RA.16}. At spin zero such structures are located 
at high excitation energies which prevents their experimental observation. 
The rotation acts as a tool to bring these exotic shapes to the yrast 
line or its vicinity so that their observation could become possible with 
future generation of $\gamma-$tracking (or similar) detectors such as 
GRETA and AGATA. Some of the studied nucleonic configurations show the 
fingerprints of clusterization and nuclear molecules. The best candidates 
for observation of extremely deformed structures, clusterization and
nuclear molecules have been identified in \cite{RA.16}. For example, the 
addition of several spin units above currently measured maximum spin of 
$16\hbar$ will inevitably trigger the transition to hyper- and megadeformed 
nuclear shapes in some of the nuclei (such as $^{36}$Ar).

  This contribution extends the investigation of \cite{RA.16} to $^{28}$Si. 
Recent experimental studies and throughout analysis of previous experimental 
data presented in Ref.\ \cite{Si28-sd.12} provide strong indications that 
the $4^+$ and $6^+$ states located at 10.946 and 12.865 MeV, respectively,
are superdeformed (SD).

\section{The details of the calculations and the results for $^{28}$Si.}

\begin{figure}[h]
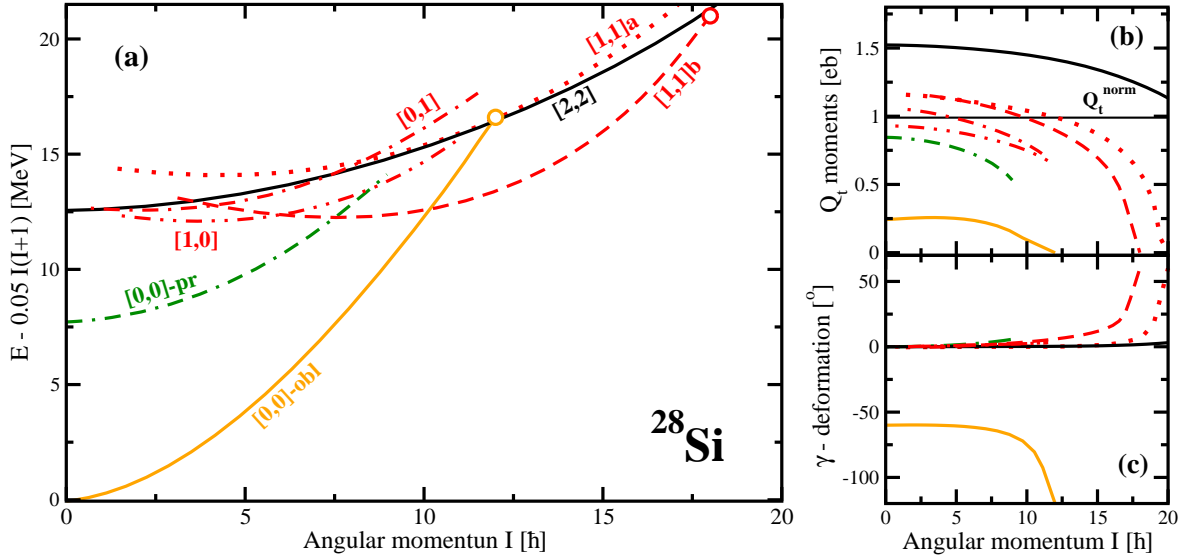

\begin{minipage}{38.0pc}
\includegraphics[width=24.5pc]{Si28_eld.eps}
\includegraphics[width=11.7pc]{Si28_gamma_new.eps}
\caption{\label{Eld} (a) The energies of the calculated configurations in $^{28}$Si 
relative to a smooth liquid drop reference $AI(I+1)$, with the inertia parameter 
$A=0.05$. This way of the presentation of the results has clear advantages as compared 
with the energy versus spin plots, see Sec.\ 4.1 in Ref.\ \cite{PhysRep-SBT} for 
details. Terminating states are encircled. (panels (b) and (c)) Calculated transition 
quadrupole moments $Q_t$ and $\gamma$-deformations of the yrast and excited SD and HD 
configurations in $^{40}$Ca. 
}
\end{minipage}
\end{figure}

  The calculations have been performed in the framework of cranked 
relativistic mean field (CRMF) theory \cite{VALR.05}. The pairing 
correlations are neglected so the calculations could be considered as 
fully realistic above $I\sim 8\hbar$. There are several reasons behind 
this neglect of pairing (see \cite{VALR.05} and recent review 
in \cite{A-rev.15}); these are the quenching of pairing by rotation 
(Coriolis anti-pairing effect), substantial shell gaps and blocking
effect. Note that only reflection symmetric shapes are considered in these 
calculations. The CRMF calculations have been performed with the NL3* 
functional \cite{NL3*} which is state-of-the-art functional for nonlinear 
meson-nucleon coupling model \cite{AARR.14} globally tested for ground state 
observables in even-even nuclei in \cite{AARR.14}. The calculated configurations 
are labeled by shorthand [$n$,$p$] labels, where $n$ ($p$) is the number of 
neutrons (protons) in the $N=3$ intruder orbitals.

\begin{figure}[h]
\begin{minipage}{34pc}
\includegraphics[width=17pc]{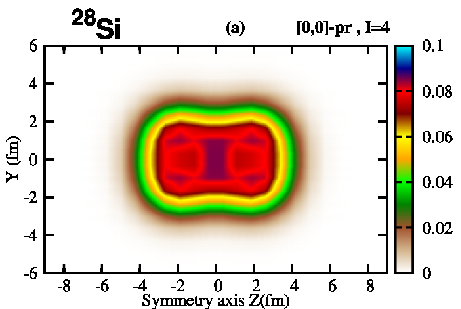}
\includegraphics[width=17pc]{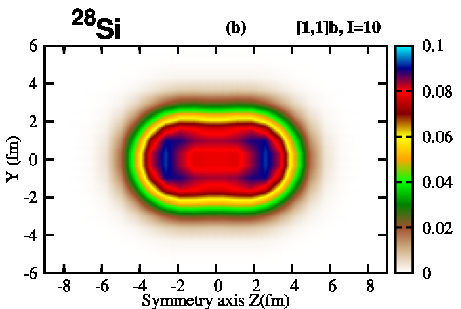}
\includegraphics[width=17pc]{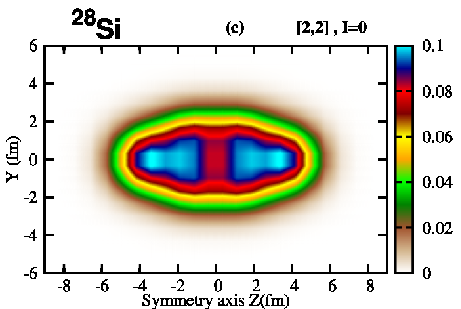}
\includegraphics[width=17pc]{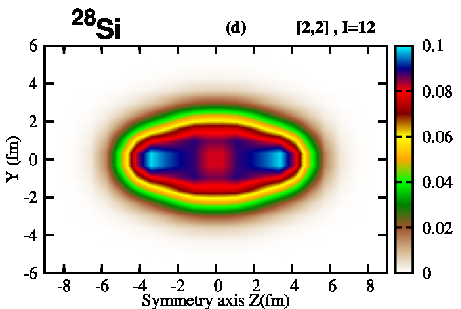}
\caption{\label{density} The self-consistent proton density $\rho_p(y,z)$ 
as a function of $y$ and $z$ coordinates for the indicated configurations 
in $^{28}$Si at specified spin values.}
\end{minipage}
\end{figure}

  Fig.\ \ref{Eld}a shows the calculated configurations in the yrast and
near yrast region. At low spin, the [0,0]-obl configuration is yrast up 
to $I\sim 10\hbar$. It has an oblate shape in agreement with experimental 
findings \cite{Si28-sd.12}. The angular momentum content of this configuration
is quite limited and it terminates at $I=12\hbar$. The yrast line at higher
spin is built by superdeformed [1,1]b configuration which terminates at
spin $I=18\hbar$. At higher spin, the hyperdeformed (HD) [2,2] configuration
becomes yrast. The transition quadrupole moments $Q_t$ and $\gamma$-deformations
of the plotted configurations are presented in Figs.\ \ref{Eld}b and c.
The density distributions of the configurations of interest are presented 
in Fig.\ \ref{density}.

   The [0,0]-pr configuration is characterized by a 
substantial quadrupole and hexadecapole deformations and shows the indications 
of the development of neck. However, present calculations do not suggest that 
it could be described as a cluster of two nuclei since the maximum of the
density is seen in the central part of nucleus (Fig.\ \ref{density}a). 
This configuration rather well describes both the excitation energies 
(compare Fig.\ \ref{Eld}a with left panel of Fig.\ 7 in \cite{TKK.09}) and the 
moments of inertia of experimental prolate band; the latter are $J^{(1)}\sim 4$ 
$\hbar^2$/MeV both in experiment \cite{TKK.09} and theory.

  Based on the comparison of experimental and calculated moments of inertia, 
the observed 4$^+$ and $6^+$ SD states are most likely associated with the 
[2,2] configuration. The calculated kinematic moment of inertia of this
configuration $J^{(1)}\sim 6.68$ $\hbar^2$/MeV, which is nearly constant for
large spin range, is only slightly above experimental one ($J^{(1)}\sim 6$ 
$\hbar^2$/MeV \cite{Si28-sd.12}). Note that there are large similarities in 
the predictions of the properties of oblate, prolate and SD bands obtained in 
present CRMF calculations and the ones within the antisymmetrized molecular
dynamics model of \cite{TKK.09}.


 Note that some indications of clusterization are present in the density 
distribution of the SD [1,1]b configuration (Fig.\ \ref{density}b). The 
HD [2,2] configuration shows clear signatures of clusterization which is 
especially pronounced at $I=0\hbar$ (Fig.\ \ref{density}c). Although the 
rotation somewhat hinders these signatures (Fig.\ \ref{density}d), they are still present at 
$I=12\hbar$.

\section{Conclusions}

  The structure of the configurations in the yrast and near yrast region
of $^{28}$Si up to $I=20\hbar$ has been studied in the CRMF theory. The 
signatures of clusterization are seen in superdeformed [1,1] configurations, 
but become especially pronounced in the hyperdeformed [2,2] configuration. 
The observation of such configurations is quite likely with future 
generation detectors such as AGATA and GRETA.
  This material is based upon work supported by the U.S. Department of Energy, 
Office of Science, Office of Nuclear Physics under Award Number DE-SC0013037.

\section*{References}
\bibliographystyle{iopart-num}
\bibliography{references14}

\end{document}